\documentclass[fleqn,usenatbib]{mnras}

\usepackage{newtxtext,newtxmath}

\usepackage[T1]{fontenc}

\DeclareRobustCommand{\VAN}[3]{#2}
\let\VANthebibliography\thebibliography
\def\thebibliography{\DeclareRobustCommand{\VAN}[3]{##3}\VANthebibliography}

\usepackage{graphicx}	
\usepackage{amsmath}	

\title[Clockwise hardness--intensity diagram in Swift~J1910.2$-$0546]{Clockwise evolution in the hardness--intensity diagram of the black hole X-ray binary Swift~J1910.2$-$0546}

\author[Saikia et al. 2023]{Payaswini Saikia,$^{1}$\thanks{E-mail: ps164@nyu.edu}
David M. Russell,$^{1}$
Saarah F. Pirbhoy,$^{1}$
M. C. Baglio,$^{1,2}$
D. M. Bramich,$^{1,3}$
\newauthor
Kevin Alabarta,$^{1}$
Fraser Lewis,$^{4,5}$
and Phil Charles$^{6}$
\\
$^{1}$Center for Astro, Particle and Planetary Physics, New York University Abu Dhabi, PO Box 129188, Abu Dhabi, UAE\\
$^{2}$INAF, Osservatorio Astronomico di Brera, Via E. Bianchi 46, I-23807 Merate (LC), Italy\\
$^{3}$Division of Engineering, New York University Abu Dhabi, PO Box 129188, Saadiyat Island, Abu Dhabi, UAE\\
$^{4}$Faulkes Telescope Project, School of Physics and Astronomy, Cardiff University, The Parade, Cardiff, CF24 3AA, Wales, UK\\
$^{5}$Astrophysics Research Institute, Liverpool John Moores University, 146 Brownlow Hill, Liverpool L3 5RF, UK\\
$^{6}$Department of Physics \& Astronomy, University of Southampton, Southampton SO17 1BJ, UK\\
}

\date{Accepted XXX. Received YYY; in original form ZZZ}

\pubyear{2023}

\begin{document}
\label{firstpage}
\pagerange{\pageref{firstpage}--\pageref{lastpage}}
\maketitle

\begin{abstract}
We present a detailed study of optical data  from the 2012 outburst of the candidate black hole X-ray binary Swift~J1910.2$-$0546 using the Faulkes Telescope and Las Cumbres Observatory (LCO). We analyse the peculiar spectral state changes of Swift~J1910.2$-$0546 in different energy bands, and characterise how the optical and UV emission correlates with the unusual spectral state evolution. Using various diagnostic tools like the optical/X-ray correlation and spectral energy distributions, we disentangle the different emission processes contributing towards the optical flux of the system. When Swift~J1910.2$-$0546 transitions to the pure hard state, we find significant optical brightening of the source along with a dramatic change in the optical colour due to the onset of a jet during the spectral state transition. For the rest of the spectral states, the optical/UV emission is mostly dominated by an X-ray irradiated disk. From our high cadence optical study, we have discovered a putative modulation. Assuming that this modulation arises from a superhump, we suggest Swift~J1910.2$-$0546 to have an orbital period of 2.25--2.47 hr, which would make it the shortest orbital period black hole X-ray binary known to date. Finally, from the state transition luminosity of the source, we find that the distance to the source is likely to be $\sim$4.5--20.8 kpc, which is also supported by the comparative position of the source in the global optical/X-ray correlation of a large sample of black hole and neutron star X-ray binaries.
\end{abstract}




\section{Introduction} \label{sec:intro}

Black hole X-ray binaries (BHXBs) are binary systems which contain a stellar-mass black hole (BH), that accretes material from a secondary, non-degenerate donor star, via an accretion disk around the central BH. BHXBs provide the easiest means to study stellar-mass BHs in detail. Gravitational waves from merging BHs \citep[e.g.][]{gw}, and isolated BHs identified from microlensing \citep[e.g.][]{sahu}, provide two complimentary methods. However, both of these are technologically challenging, the events are short-lived, and their locations are unpredictable. More than 50 years of studies of tens of BHXBs have led to knowledge of their mass distribution \citep[e.g.][]{farr,ozel,jon21}, constraints on their spins \citep[e.g.][]{spin2,spin1}, and how they affect matter in their close proximity \citep[e.g.][]{Lasota2001,Fender01}. Learning about these objects also enables us to uncover the evolutionary end point of high-mass stars, test General Relativity in extreme gravitational fields, and study their feedback in terms of the radiation and the outflows that they produce.

The majority of BHXBs are transient systems that spend most of their time in quiescence, experiencing very low levels of accretion. However, occasionally they undergo outbursts, where the accretion rate and the luminosity increases by several orders of magnitude. During an outburst, several spectral states can be identified \citep[see e.g.][]{Homan2005,mcclintock2003black}, mainly the hard state when the X-ray spectrum is dominated by a hard X-ray component that can be described by a power law and the soft state where the X-ray emission is dominated by a thermal component and a weak power-law tail. There are also intermediate states during the transition between the hard and soft states, when both the power law and the thermal component contribute. BHXBs usually transit through these different spectral states producing a ‘q-shaped’ hysteresis loop, moving counterclockwise in the hardness--intensity diagram \citep[HID,][]{my,homanhid,f2009,b2010}. During an outburst, compact jets are often ejected in the hard state, and ballistic jets are launched when the source transitions from the intermediate to the soft state \citep{f2009}. There is a strong observational connection between accretion and ejection during the hard state, as shown by a correlation between the radio and X-ray emission of BHXBs \citep[e.g.][]{gallofenderpooley2003,corbel2003,Corbel2013,esp,gallo2008rx}, which can also be extended to supermassive BHs by taking into account the mass of the BH \citep[e.g.][]{merloni2003,Falcke2004,saikia2015,saikia2018}.

Swift~J1910.2$-$0546 (also known as MAXI~J1910$-$057) is a BHXB candidate discovered by the Neil Gehrels Swift Observatory \citep[\emph{Swift};][]{burrows} and the Monitor of All-sky X-ray Image \citep[MAXI;][]{maxi} as a new X-ray transient, when it went into an outburst in 2012 May \citep{Krimm2012,Usui2012}. Soon after, its optical/near-infrared (NIR) counterpart was discovered \citep{Rau2012}. A search for its radio counterpart initially failed to yield a detection \citep{Fogasy2012}, and finally was detected in 2012 August \citep{King2012ATel4295}. Its X-ray spectral and timing properties \citep{Nakahira14,Degenaar2014}, and optical spectroscopy \citep{Charles2012,Casares2012} revealed its nature as a low mass X-ray binary (LMXB), likely a black hole candidate that underwent hard-to-soft X-ray state transitions. The source decayed slowly and performed several state transitions, before returning to the hard state in 2012 November \citep{Nakahira14,Degenaar2014}. After this, the source showed some re-brightening activities in 2013 and 2014 \citep{Tomsick2013ATel5063,paper2}, and a new outburst in 2022 \citep{2022xray,2022opt1,2022opt2,2022radio,paper2}.

Almost all of our knowledge about Swift~J1910.2$-$0546 has been gained from its first outburst in 2012. Many properties of the system, including its distance ($d$), orbital period ($P_{orb}$), black hole mass ($M_{\rm BH}$), companion star mass ($M_{\rm *}$) and spectral type, and orbital inclination ($i$), remain largely unknown. The only constraints reported to date are fairly loose. \cite{Nakahira14} provide limits for the system parameters as $M_{\rm BH} > 2.9 M_{\odot}$, $d > 1.7$ kpc and $i\lesssim 60^{\circ}$. The orbital period of the source was initially thought to be relatively short ($\sim$2-4 hrs) based on the highly variable optical emission \citep{llyod}, but spectroscopic studies later suggested it to be greater than $\sim 6.2$ hrs \citep{Casares2012}. The spin of the black hole may be unusually retrograde, but this conclusion depends on the truncation of the inner disk in the intermediate spectral state \citep{Reis2013}.

Here we present a detailed multi-wavelength study of the 2012 outburst, mainly focusing on new optical data with the Faulkes telescopes (four different filters), along with previously unpublished observations in the UV wavelengths (five different filters). We focus on understanding the complex spectral evolution of the source by performing a energy-dependent analysis of the HID and placing them in the context of our optical and UV analysis. In addition, we use the optical/UV spectral energy distribution and multi-wavelength correlations to investigate the physical mechanisms contributing to the emission in these wavebands. We also present a high cadence optical analysis to put constrains on the orbital period of the system, and compare the quasi-simultaneous optical/X-ray data of the source with a compilation of data from other low-mass BHXBs to constrain its distance. The data used in the study are presented in Section 2. We discuss the results in Section 3, including a detailed analysis of the dip in the intensity almost 90 days into the outburst, the peculiar HID and the optical behavior of the source during transition, high cadence optical observations, optical/X-ray correlations, spectral energy distributions, and implications on the distance to the source. Finally we present the conclusion of the study in Section 4.

\section{Observations} \label{sec:obs}

\subsection{Faulkes Telescope / LCO monitoring} \label{sec:obsLCO}

We started monitoring Swift~J1910.2$-$0546 with the two Faulkes Telescopes on 2012 June 14th (MJD 56092), a few weeks after the initial discovery of the source. Imaging was carried out in the Bessell $B$, $V$, $R$ and SDSS $i^{'}$ filters with the 2-m Faulkes Telescopes at Haleakala Observatory (Maui, Hawai`i, USA) and Siding Spring Observatory (Australia), as well as the 1-m telescopes at Siding Spring Observatory (Australia), Cerro Tololo Inter-American Observatory (Chile), McDonald Observatory (Texas), Teide Observatory (Tenerife) and the South African Astronomical Observatory (SAAO, South Africa). These are robotic telescopes optimized for research and education \citep[e.g.][]{Lewis2018} and are part of the global network, Las Cumbres Observatory \citep[LCO;][]{Brown2013}. These observations are part of an on-going monitoring campaign of $\sim$50 LMXBs \citep{Lewis2008} co-ordinated by the Faulkes Telescope Project.

Aperture photometry was performed by the ``X-ray Binary New Early Warning System (XB-NEWS)'' data analysis pipeline \citep{Russell2019,Pirbhoy2020,Goodwin2020}. The pipeline downloads images and calibration data of all targets of interest from the LCO archive, performs several quality control steps to ensure a good quality of the images, computes an astrometric solution on each image using Gaia DR2\footnote{\url{https://www.cosmos.esa.int/web/gaia/dr2}} positions, and then performs aperture photometry of all the stars. The method described in \citet{Bramich2012} is used to solve for zero-point magnitude offsets between epochs, and flux calibration of all the stars is achieved using the ATLAS All-Sky Stellar Reference Catalog \citep[ATLAS-REFCAT2,][]{Tonry2018}\footnote{\url{https://archive.stsci.edu/prepds/atlas-refcat2/}}, which includes PanSTARRS DR1, APASS, and other catalogues to extract the magnitudes of the source. When the source is not detected above the detection threshold by the pipeline, XB-NEWS performs forced photometry at the known location of the source. Magnitudes with errors $> 0.25$ mag are excluded as these are either very marginal detections, or non-detections.

During the 2012 outburst, we have magnitude measurements of Swift~J1910.2$-$0546 from a total of 72, 44, 37 and 29 images in the $B$, $V$, $R$, and $i^{'}$-bands, respectively, between 2012 June 14th (MJD 56092) and 2013 Jan 31st (MJD 56323). Most of these are from regular monitoring, with images taken every few days in $B$, $V$, $R$ and $i^{'}$-band. In addition, we have 23, 42 and 43 images taken in $B$-band on 2012 July 16th, 19th, 24th to test for short-term, periodic or aperiodic variability of the source. 

\subsection{X-ray monitoring} \label{sec:obsXray}

Swift~J1910.2$-$0546 was monitored by the X-Ray Telescope \citep[XRT;][]{burrows} onboard \emph{Swift} every few days over its 2012 outburst. The source was observed in the Windowed Timing mode \citep[WT,][]{hill} until 2012 November, with an average exposure time of a few kiloseconds in each observation. We used all observations during the outburst with target IDs 00032480 (16 observations, 2012 June-August), 00529076 (2 observations, 2012 July) and 00032521 (49 observations, 2012 August-November). Due to the Sun constraint, no observations were taken by \emph{Swift}/XRT from 2012 November until 2013 March. We retrieved the X-ray light curve of Swift~J1910.2$-$0546 from all these observations using the online \emph{Swift}/XRT data products generator\footnote{\url{https://www.swift.ac.uk/user objects/}} maintained by the \emph{Swift} data center at the University of Leicester \citep[see][]{evans1,evans2}. We binned the data by observation to extract the 2-10 keV light curve, as well as the 1.5-10 keV/0.6-1.5 keV X-ray hardness ratio for our analysis. The X-ray fluxes were obtained using a photon index of
$\Gamma\sim$1.7, and a hydrogen column density value of $N_H=(3.5\pm0.1)\times 10^{21}\, \rm cm^{-2}$ \citep{Degenaar2014}. We note that since we convert from count rates to fluxes in the same energy range, a slightly different value of photon index would make very little difference to the resulting fluxes, and source evolution.

The Burst Alert Telescope \citep[BAT;][]{krimm} on-board the \emph{Swift} observatory is an all-sky monitor to search for gamma-ray bursts, which also monitors the hard X-ray sources. We gathered the light curve of Swift~J1910.2$-$0546 in the 15–50 keV range with a one-day time bin, from the \emph{Swift}/BAT transient monitor archive\footnote{\url{https://swift.gsfc.nasa.gov/results/transients}} provided by the \emph{Swift}/BAT team \citep[][]{krimm}. We also acquired the MAXI light curve of the source obtained with the Gas-Slit Camera \citep[GSC;][]{gsc} from the MAXI public web page\footnote{\url{http://www.maxi.riken.jp}} in the 2-20 keV range during the outburst.

\subsection{Ultraviolet monitoring} \label{sec:obs<UVOT}

We gathered publicly available \emph{Swift} Ultraviolet and Optical Telescope
\citep[UVOT;][]{Roming} observations from the NASA/HEASARC data center, in all available filters with the corresponding central wavelengths \citep{poole} : $uvw2$ (1928 Å), $uvm2$ (2246 Å), $uvw1$ (2600 Å), $u$ (3465 Å), $b$ (4392 Å), and $v$ (5468 Å). We used all data taken in the 2012 outburst with target IDs 32480, 32521, 32742, 524642 and 529076. We used the pipeline processed images and obtained the magnitude of the source using the {\tt uvotsource} HEASOFT routine. To extract the magnitudes and flux densities, we used a standard aperture of 5 arcsec centered on the source, and an empty region of 30 arcsec radius close to the source as the background region. The source was detected in almost all the observations by \emph{Swift}/UVOT. For our analysis, we select only those pointings where the significance of the detection above the sky background is higher than 5$\sigma$.

\begin{figure*}
\centering
\includegraphics[height=21.5cm,angle=0]{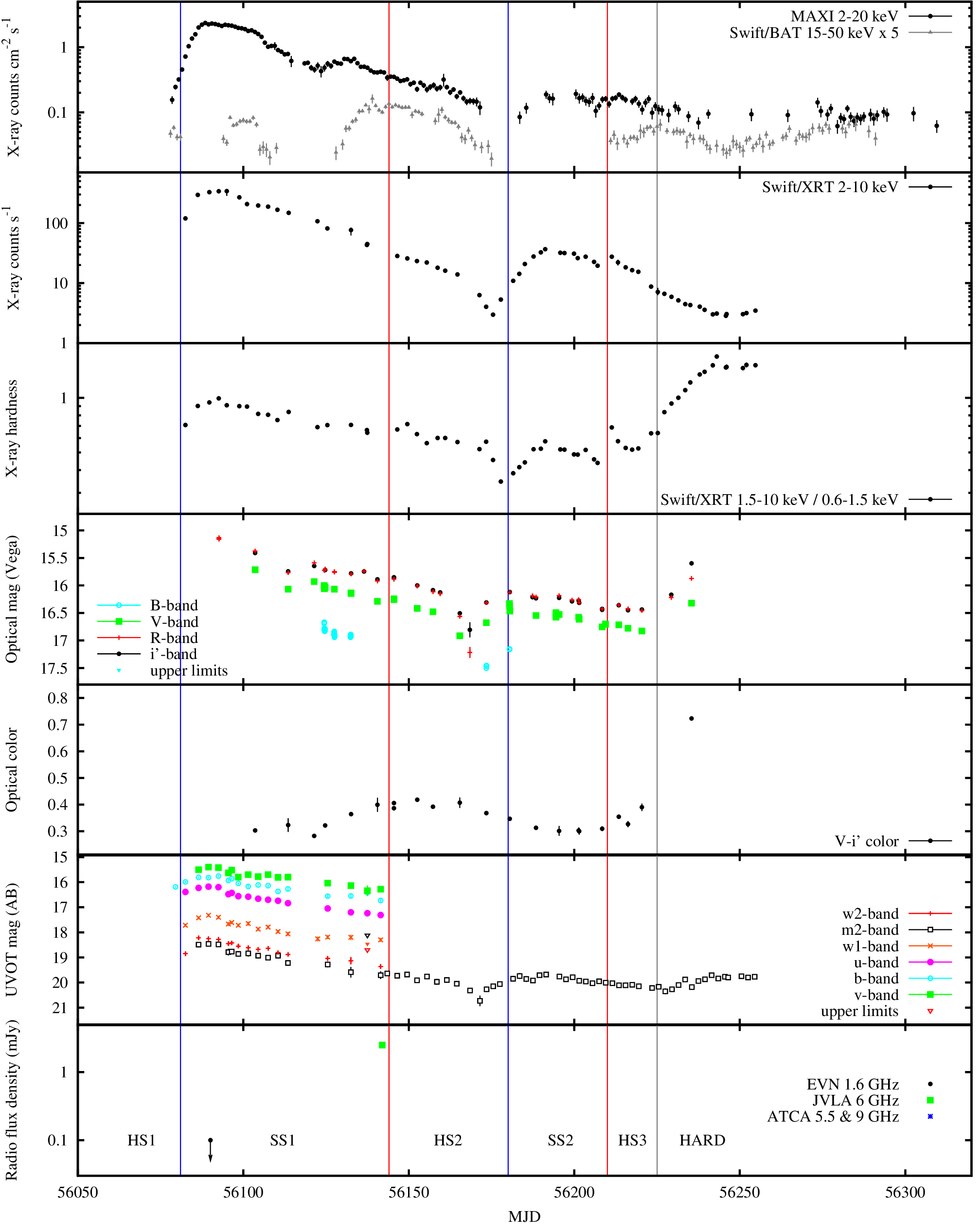}
\caption{Multi-wavelength light curves of Swift~J1910.2$-$0546 from 2012 to 2015, with the blue vertical lines marking the transitions from the hard to soft states, red lines from soft to hard states, and the grey line showing the transition to a pure hard state on 2012 Oct 25th (MJD 56225) in the 2012 outburst. (a) MAXI 2--20 keV and \emph{Swift}/BAT 15--50 keV X-ray count rates; (b) \emph{Swift}/XRT 2--10 keV X-ray count rate; (c) X-ray hardness from \emph{Swift}/XRT; (d) Optical magnitude in $B$, $V$, $R$ and ${i}^{\prime }$-bands; (e) Optical color ($V$-${i}^{\prime }$); (f) Optical--UV magnitude in the six \emph{Swift}/UVOT filters; (g) Radio flux density from the literature (see text).} 
\label{fig:lc-large}
\end{figure*}

\subsection{Radio data} \label{sec:obsRadio}

All radio observations of Swift~J1910.2$-$0546 are taken from the literature. Initially, the European VLBI Network (EVN) did not detect the source (flux density $< 0.1$ mJy beam$^{-1}$ at 1.6 GHz) near the start of the outburst, in 2012 June \citep{Fogasy2012}. Then, a relatively bright radio detection of `nearly 2.5 mJy' (the exact value and error are not provided) was made by the Karl G. Jansky Very Large Array \citep[VLA;][]{vla} at 6 GHz in 2012 August \citep{King2012ATel4295}.

\section{Results and Discussion} \label{sec:results}

The multi-wavelength evolution of Swift~J1910.2$-$0546 from the start of the 2012 outburst is shown in Fig. \ref{fig:lc-large}. We present the time evolution of the brightness and the hardness and color at different wavelengths (Fig. \ref{fig:lc-large}).

In Fig. \ref{fig:lc-large}(a) the MAXI and \emph{Swift}/BAT light curves are shown (daily averages; only detections with $\geq 4 \sigma$ significance are plotted). The MAXI light curve can be described by a fast rise, slow decay profile, with reflares during the decay and a plateau. The clear difference in trends between the two X-ray energy ranges obtained with MAXI (2--20 keV) and \emph{Swift}/BAT (15--50 keV) evolution during 2012 highlight the spectral state changes that are occurring during the outburst. Panel (b) in Fig. \ref{fig:lc-large} shows the \emph{Swift}/XRT light curve. The 2--10 keV light curve is similar in profile to the MAXI light curve, but shows a late reflare more clearly. After the \emph{Swift}/XRT monitoring ended on 2012 Nov 23rd (MJD 56254), MAXI and \emph{Swift}/BAT continued to detect the source for $\sim$40 d before fading below the detection limits. Panel (c) represents the X-ray hardness ratio of the source, where it first softened, and then hardened. 

In Fig. \ref{fig:lc-large}(d) we present the LCO optical data in $B$, $V$, $R$ and ${i}^{\prime }$-bands. After the transition to the pure hard state, there is a gap in the coverage due to a Sun constraint. We plot the optical $V$-${i}^{\prime }$ color in Fig. \ref{fig:lc-large}(e), which shows a significant color change during the transition to the pure hard state, as the source became redder (see Section \ref{sec:optintransition}). Panels (f) and (g) show the \emph{Swift} UVOT optical--UV, and radio light curves, respectively. All six UVOT filters were used for the first half of the 2012 outburst, after which only the $uvm2$-band filter was used.

\begin{figure}
\centering
\includegraphics[width=8.7cm,angle=0]{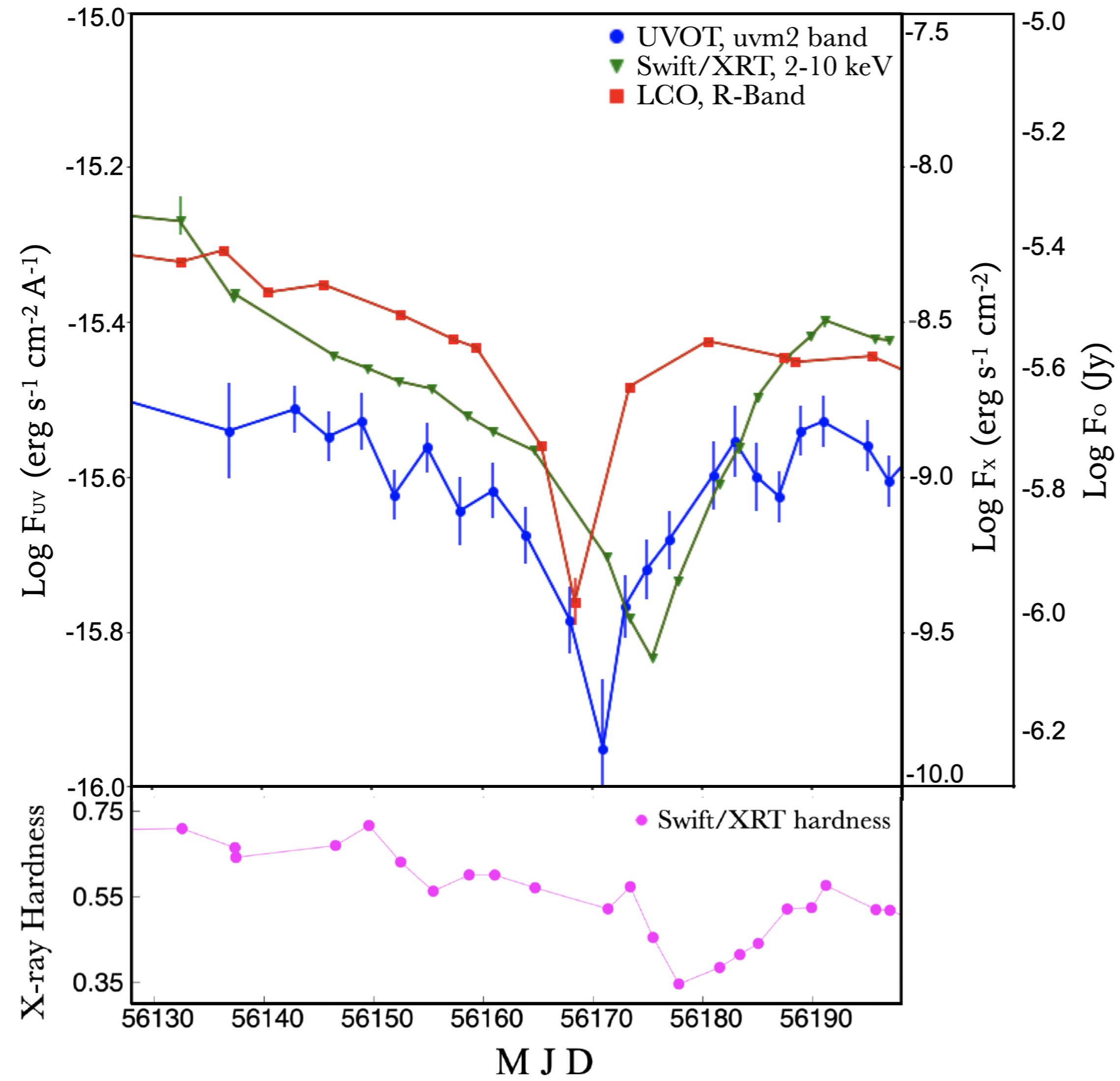}
\caption{The upper panel shows the delay in the dip as measured in the optical (LCO $R$-band, in red squares), UV (\emph{Swift} UVOT $uvm2$, in blue filled circles) and X-ray (\emph{Swift}/XRT, in green inverted triangles) wavebands. The lower panel shows the X-ray hardness (1.5-10 keV/ 0.6-1.5 keV) with \emph{Swift}/XRT.} 
\label{fig:dip}
\end{figure}

\subsection{Dipping event} \label{sec:dip}

We observe a prominent dip in the optical intensity around 90 days into the outburst, which happened a few days before the source entered the soft state (see Fig. 2). The observed optical dip occurs slightly before the UV and X-ray dips reported in the literature \citep[][]{Nakahira14,Degenaar2014}. The dip is prominent in all the wavebands, with the optical magnitudes in the $R$-band fading by 0.65 mag (a factor of $\sim$1.9), the UV magnitude in the $uvm2$ band by $\sim$0.8 mag (a factor of $\sim$2.1), and a factor of $\sim$5.4 in the X-ray count rate at 2-10 keV energy range. After the dip, the fluxes at all wavelengths returned back to the same level as they were before the dip occurred.

The timing of the dip is not coincident in the three wavelengths, nor is it repeated, and hence it is unlikely to be due to an eclipse. The time delay observed between the different wavelengths for the occurence of the dip is illustrated in Fig. 2.  We find that the optical minimum occurred first ($\sim$90 days after the onset of the outburst; 2012 Aug 29th, MJD 56168), followed by the the UV minimum which was delayed from the optical by three days ($\sim$93 days; Sep 1st, MJD 56171). The X-ray minimum happened last, $\sim$97.5 days after the outburst started (Sep 5th, MJD 56175.5). The observed delay in the occurence of the dip at different wavelengths can be attributed to radial inflow of mass instability from the outer edge of the accretion disk towards the inner region. The delay time reflects the viscous timescale of the disk \citep{Degenaar2014}, as the accreting matter propagates from the outer part of the accretion disk emitting in optical and UV wavelengths to the inner accretion disk that emits in X-rays \citep{Nakahira14}. This X-ray dip was closely followed by the transition to the soft state (Sep 7th, MJD 56177.8), which suggests that there is a link between the matter propagating through the accretion disk and the spectral state change. As noted by \cite{Degenaar2014}, this could also be explained by the collapse of a hot inner flow before the source transitions to a soft state \citep{veledina2011}. However following such a collapse, a recovery of the hot inner flow is expected causing the light curves to peak in the UV emission followed by the infrared emission as the source transitions back to hard state \citep{veledina2013}, but that is not observed in the case of Swift~J1910.2$-$0546 \citep[for a detailed discussion, see][]{Degenaar2014}.

\subsection{Hardness--intensity diagram} \label{sec:hid}

\begin{figure*}
\centering
\includegraphics[width=17.4cm,angle=0]{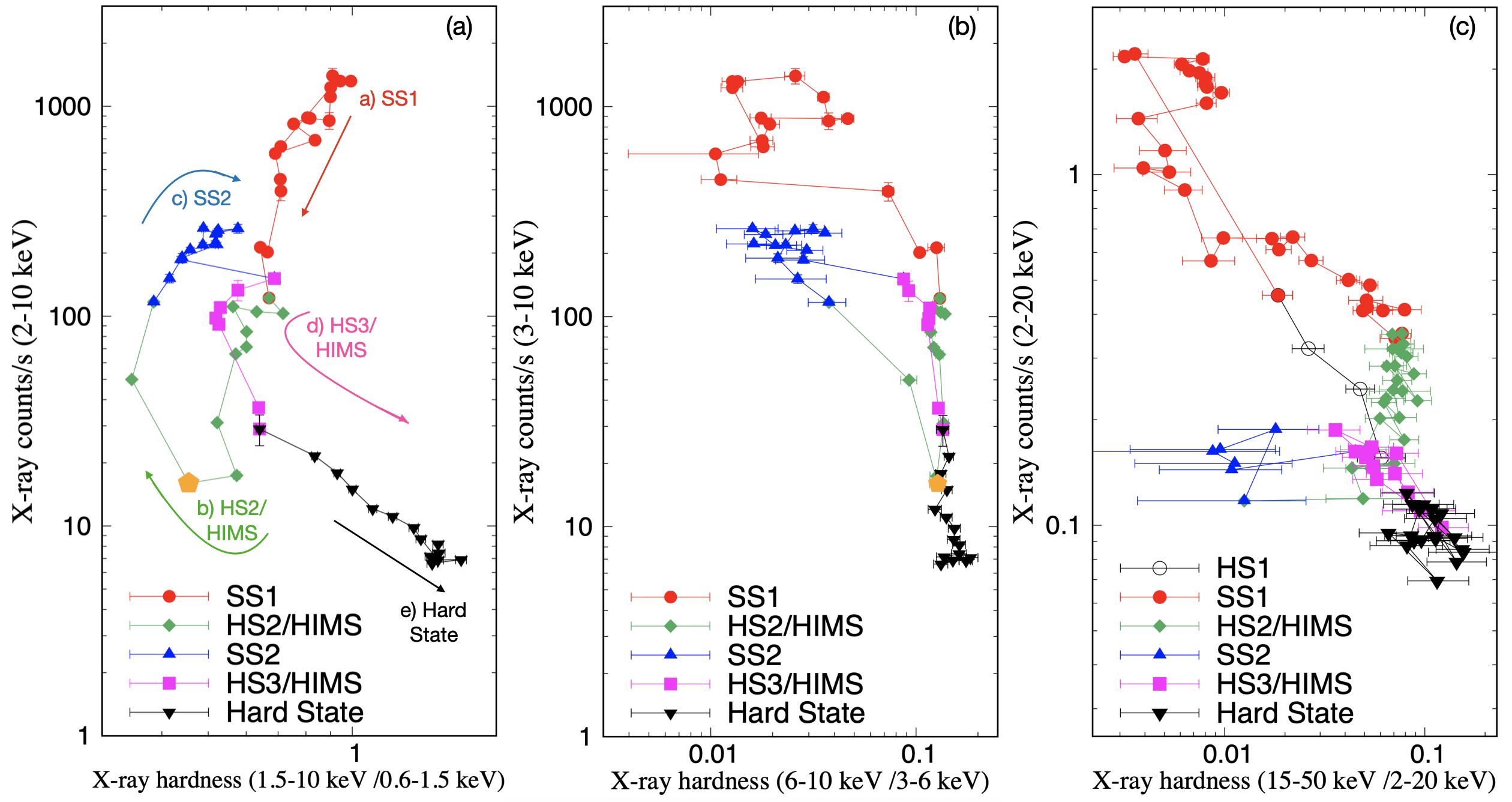}
\caption{X-ray hardness--intensity diagrams of Swift~J1910.2$-$0546 using (a) Swift/XRT count rates at 2-10 keV, 1.5-10 keV and 0.6-1.5 keV energy ranges (with arrows overplotted to show the evolution of the HID during the outburst), (b) Swift/XRT count rates at 3-10, 6-10 and 3-6 keV energy ranges, (c) MAXI/GSC 2-20 keV and Swift/BAT 15-50 keV energy ranges. Different colors and symbols are used to represent the various states and stages of the outburst (see text). The orange pentagon in the first two panels represents the minimum of the X-ray dip on 2012 Sep 5th (MJD 56175.5, see Section 3.1).}
\label{fig:HIDs}
\end{figure*}

We investigated the X-ray spectral evolution of Swift~J1910.2$-$0546 during the outburst using its HID to study its various spectral states (see Fig. \ref{fig:HIDs}). The two previous studies that have analysed the HID of Swift~J1910.2$-$0546 using two different X-ray energy ranges, report contradicting results. \cite{Reis2013} used \emph{Swift}/XRT pointing observations and the hardness ratio (2-10 keV/0.6-2 keV) calculated from unabsorbed fluxes (obtained using a simple model consisting of an absorbed disk blackbody plus power law) and found that the source followed the typical ‘q-shaped’ pattern when plotted against a sample of BHXBs. However, \cite{Nakahira14} analysed the evolution of the HID using a hardness ratio of 3–5 keV/5–10 keV from \emph{Swift}/XRT, and X-ray luminosity obtained with model fits, and reported two unusual features in the HID. They found that the source exhibited hard-to-soft and soft-to-hard spectral state transitions twice; and the soft-to-hard transition in the second case occurred at a higher luminosity than the hard-to-soft transition.

In order to investigate and explain the unique properties of the HID of Swift~J1910.2$-$0546 observed by previous studies, we conduct the first detailed study of the energy dependence of the HID using different soft and hard X-ray energy ranges. We use the soft X-ray \emph{Swift}/XRT data in the 2-10 keV, 1.5-10 keV and 0.6-1.5 keV ranges, as well as the MAXI/GSC data in the 2-20 keV range and hard X-ray \emph{Swift}/BAT data in the 15-50 keV range. For this analysis, we do not perform any spectral fitting of the data to extract the fluxes, but simply use the count-rates and hardness ratios which are model-independent. We adopt the hardness ratio intervals labelled by \cite{Nakahira14}, who divided the entire outburst period of the source into five intervals of hard state (HS) and soft state (SS) starting from 2012 May 31st (MJD 56078) as the zeroth day. They divided the different periods as HS1 (t$<$ 3 days), SS1 (3 days $<$ t $<$ 66 days), HS2 (66 days $<$ t $<$  102 days), SS2 (102 days $<$ t $<$ 132 days), and HS3 (t $>$ 132 days). They noted that HS2 could also be characterized as the hard--intermediate state (HIMS). \\cite{Degenaar2014} noted that the X-ray spectrum is completely dominated by the hard power law component only after MJD 56225 (see also Fig. 1c, where the hardness ratio increases around MJD 56233). Hence, we label the last phase of the outburst (t $>$ 147 days) as the pure hard state.

In Fig. \ref{fig:HIDs}a, we show the 2-10 keV count rates vs. the 1.5–10/0.6–1.5 keV hardness ratio from \emph{Swift}/XRT observations, where the hardness ratio roughly represents the ratio of the power law over the disk emission. In Fig. \ref{fig:HIDs}b, we plot the \emph{Swift}/XRT 3-10 keV count rates vs. the 6–10/3-6 keV hardness ratio. Finally in Fig. \ref{fig:HIDs}c, we use a much harder ratio and plot the 2-20 keV count rates vs. the 15–50/2–10 keV hardness ratio from \emph{Swift}/BAT and MAXI/GSC, where the hardness ratio purely traces the hardness of the power law up to 50 keV with a negligible disk contribution. In all the HIDs, we find a complex evolutionary track, with a clockwise loop during the second transition (highlighted in Fig. \ref{fig:HIDs}a with arrows), where the transition to the soft state (HS2/HIMS to SS2) occurs at a fainter flux than the transition out of the soft state \citep[SS2 to HS3/HIMS, as also reported in][]{Nakahira14}. This is highly unusual, and is opposite to the typical, well known hysteresis in the HIDs of BHXB transients \citep{homanhid,hb,dunn2010}. In Fig. \ref{fig:HIDs}c, we discover that there is evidence of a clockwise loop of the power law component half way down, as the HS1 rises softer than the SS1 fades (this is not observed in the previous two panels, as we do not have data in HS1 from \emph{Swift}/XRT). However the additional clockwise loop just after the X-ray dip (prominent in the previous two panels where the disk substantially contributes towards the X-rays), is not evident in Fig. \ref{fig:HIDs}c. Hence, it is possible that the clockwise loop in Fig. \ref{fig:HIDs}a arises due to brightening of the disk by $\sim$1 order of magnitude, while the power law component remains almost constant. In this scenario, the dip and recovery could represent a decrease, and then an increase, in the mass accretion rate propagating through the disk. When the increase in mass reaches the inner regions it causes a higher mass accretion rate in the inner disk, increasing the soft X-ray flux \citep[a higher disk temperature is also observed at this time;][]{Degenaar2014}.

Such complex X-ray spectral changes featuring unusual loops have also been observed in MAXI~J1820+070, just after the appearance of a large modulation in the optical light curve \citep{thomas1820}. The two other BHXBs showing clear clockwise evolution in the HID are MAXI~J1535$-$571 and 4U~1630$-$47. MAXI~J1535$-$571 followed the more common counter-clockwise trajectory during the outburst, but showed a clockwise evolution in the HID during the first flaring event \citep{cuneo,Parikh2019}. 4U~1630$-$47 also followed a counter-clockwise track during its 1998 outburst, but often moved in a clockwise direction during the 2002–2004 \citep{Tomsick2002outburst} and the 2010 outburst \citep{Tomsick2010outburst}, in both of which the system stayed at a lower hardness level compared to its 1998 outburst. \cite{Tomsick2010outburst} noted that one main difference between the 1998 and the 2002 or 2010 outbursts, is the presence of radio emission coming from the jet during the 1998 outburst, which was absent in the later ones. For Swift~J1910.2$-$0546, there is a bright radio detection of nearly 2.5 mJy at 6 GHz near the end of SS1 before the clockwise loop \citep{King2012ATel4295}, so it is probably a discrete ejection launched from the first entry into SS1 and hence due to the lack of data, we cannot comment on the jet affecting (or not affecting) the loop in this source. However, because the loop is more prominent in Fig. \ref{fig:HIDs}a and Fig. \ref{fig:HIDs}b with the soft X-rays, and it happened just a few days after the dip in the intensity (linking the changes in matter propagating in the disk to the change in the spectral state of the source), we suggest that the secondary clock-wise loop is caused by a brightening of the disk (making it softer and brighter before decaying) while the power law remains relatively unchanged.

\begin{figure*}
    \centering
    \includegraphics[scale=0.28]{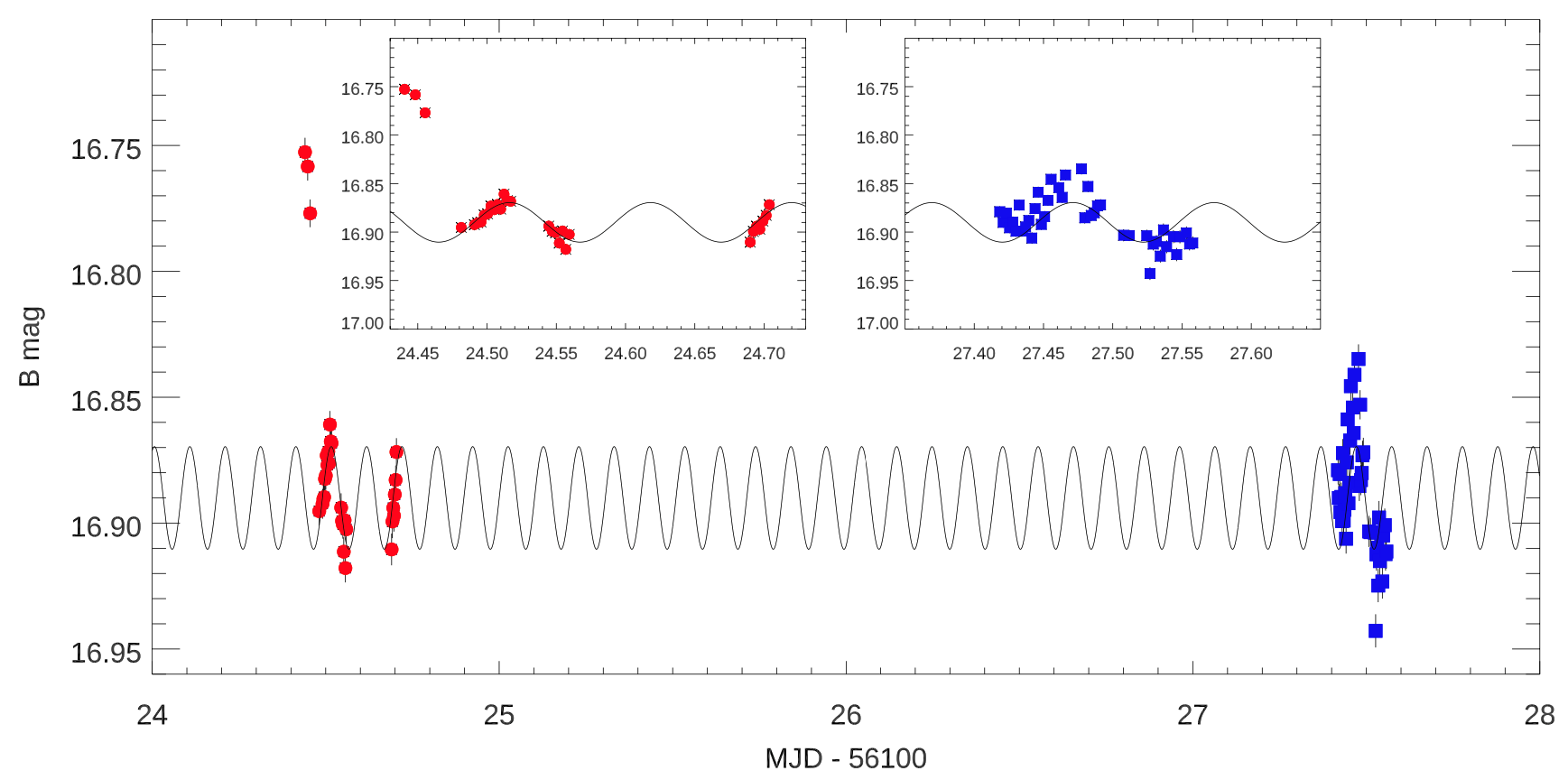}
    \caption{$B$-band light curves of Swift~J1910.2$-$0546 in two epochs : July 16th (MJD 56124, red circles) and July 19th (MJD 56127, blue squares). The two insets show the zoom of the two light curves, separately. Superimposed, the sinusoidal fit of the light curves together (black solid line). Errors are indicated at the 68$\%$ confidence level. Magnitudes are not de-reddened.}
    \label{fig:fast_variability}
\end{figure*}

\subsection{Optical behavior during the X-ray transition} \label{sec:optintransition}

During the transition of the source from a soft/intermediate to the pure hard state at the end of the outburst, around 2012 Oct 25th ($\sim$MJD 56225), we discover an increase in the optical and UV flux, as the X-rays continuously decay (by amplitudes of 0.6 and 0.3 in ${i}^{\prime }$ and $uvm2$ bands, respectively, by MJD 56235). The flux increase is more prominent at longer wavelengths, with the $V$-${i}^{\prime }$ colour changing from $\sim$0.3-0.4 to 0.73, which is consistent with the IR brightening and IR colour change reported in \cite{Degenaar2014} a few days before (MJD $\sim$56215).

Similar to Swift~J1910.2$-$0546, a detectable optical and NIR excess (above the disk component) has also been observed during the onset of a jet in the hard spectral state of many other BHXBs \citep[see][and the references within]{Kalemci2013,saikia2019}. Although a change of UV emission during state transitions is not as common, there are examples of sources like GX~339-4 where a UV flux dip associated with jet quenching was reported simultaneously with an emission dip in optical/NIR wavelengths when the source transitioned from a hard to soft state in its 2010 outburst \citep{YanYu2012}, and MAXI~J1820+070 where a UV re-brightening was detected during the decay of the 2018/2019 outburst accompanied by an increase in the X-ray hardness \citep{1820uv}.

Another interpretation for the same could be the presence of a hot inner flow \citep{veledina2011,veledina2013}, which can also explain the time delays between the dipping events at different wavelengths (see Section \ref{sec:dip}) and the subsequent transition to the soft state. However the collapse and recovery of a hot inner flow requires that the rise in the UV emission occurs before the rise in NIR wavelengths, which is not the case for Swift~J1910.2$-$0546, where the NIR rise happened $\sim$20 days before the rise in UV emission \cite[see also][]{Degenaar2014}. The jet interpretation, on the other hand, can explain the observed delay sequence in the rise of the emission at different wavelengths where the optical and UV rise follows the NIR excess \citep{Russell2013a} - with the rise of the NIR excess around 2012 Oct 15th (MJD 56215), followed by optical in Oct 29th (MJD 56229) and UV in Nov 4th (MJD 56235). Generally in such a case, a delay of 5--15 days is expected between the NIR peak and the transition from the soft to the hard state \citep{Kalemci2013}, but due to lack of coverage in NIR, the peak of the secondary excess cannot be constrained for the case of Swift~J1910.2$-$0546. However from the observed delay, we attribute this increase in the optical/NIR and UV emission at the same time as the transition to the hard state to the revival of a jet revealing itself in these wavelengths \citep[eg.][]{Kalemci2013,saikia2019}. This is also supported by the color-magnitude diagram where the data during the transition shows significant dispersion from the disk model, suggesting that the synchrotron jet significantly contributes towards the emission at optical wavelengths during the transition \citep{paper2}. It shows that the de-reddened $V-{i}^{\prime}$ spectral index is $-0.7$ and the $uvm2-{V}$ spectral index is $-1.3$ for this brightening. This is too steep for a hot flow, which is expected to be almost flat ($\alpha\sim$0, or $\alpha>0$ from the outer regions of the hot flow), but is typical of optically thin jet synchrotron emission rising towards the NIR.

\subsection{High cadence optical observations}\label{Sec:fast_optical_variability}

During the soft state (SS1), high cadence optical observations (probing approximately the frequency range $10^{-4}-10^{-3}$ Hz) were taken in $B$-band with the LCO 2-m telescopes in three epochs: July 16th (MJD 56124, 23 exposures, for a total time on source of $\sim2.8$hrs and a time resolution of $\sim7.8$mins)
, July 19th (MJD 56127, 42 exposures, for a total time on source of $\sim3.3$hrs and a time resolution of $\sim4.5$mins)
and July 24th (MJD 56132, 43 exposures, for a total time on source of $\sim2.7$hrs and a time resolution of $\sim4$mins).
For the first two light curves, significant variability is detected, with a $4.31\pm0.11\%$ and $2.11\pm0.09\%$ fractional rms deviation in the flux \citep[corrected for white noise due to estimated measurement uncertainties, calculated following the method described in][]{vaughan,Gandhi2010}, respectively. The third epoch instead presents an almost flat curve, with lower variability (the fractional rms for the light curve is $0.77\pm0.11\%$). We looked for possible sinusoidal modulations in the first two light curves. The results of the analysis are shown in Fig. \ref{fig:fast_variability}.

We considered the two variable light curves but excluding the first three observations taken on July 16th (MJD 56124) that resemble a flare-like event. We found that the remaining points are sinusoidally modulated at a period of $2.36\pm0.01$ hours and a semi-amplitude of the oscillation of $0.02\pm0.01$ mag ($\chi^2$/dof=494.5/61). The sinusoidal fit is preferable to the fit with a constant brightness at $5\sigma$ confidence level according to an $F$-test. However, we note that with our observations we see this sinusoidal modulation in only two and half cycles, and a follow-up is required to confirm that there is a period, as opposed to accretion activity that is not periodic.

The observed periodicity is similar to that detected for the same system and outburst according to observations carried out at the Nordic Optical Telescope \citep[NOT, Roque de los Muchachos Observatory, La Palma;][]{nordic} on July 21st (MJD 56129), reported by \citet{Casares2012}. In particular, they detected a modulation of amplitude $\sim0.02$ mags on a $\sim4$hr time scale.
Such a periodicity may be linked to the presence of a hot spot on the surface of the outer disk. Under this assumption, we can derive an estimate of the orbital period of the source by approximating the outer radius $R_{\rm out}$ of the disk with its tidal radius $R_{\rm tid}$. In particular, from \citet{egg,King1996}, $R_{\rm out}\sim R_{\rm tid}=0.9R_{\rm L}$, where $R_{L}$ is the Roche lobe radius of the BH, given by:

\begin{equation}
\frac{R_{L}}{a}=\frac{0.49\,q^{2/3}}{0.6\,q^{2/3}+\rm ln(1+q^{1/3})},
\end{equation}
where $q$ is the ratio between the BH mass and the companion star mass ($M_{\rm BH}$ and $M_{*}$, respectively), and $a$ is the binary separation (that we can consider as an upper limit to the radius of the companion star orbit).
We can then use the third law of Kepler:

\begin{equation}
\frac{P_{\rm out}}{P_{\rm orb}}\propto \left( \frac{R_{\rm out}}{a}\right)^{3/2},
\end{equation}
where $P_{\rm orb}$ and $P_{\rm out}$ are the orbital period of the companion star and that of the outer regions of the disk, respectively.
From \citet{Nakahira14}, we know a lower limit to the BH mass ($M_{\rm BH}>2.9\, M_{\odot}$); to get the most conservative constraint on the orbital period, we use the maximum possible mass of an $M$-type star \citep[as derived from the fit of quiescent spectral energy distribution of the system;][]{paper2}, i.e. $\sim 0.6\,M_{\odot}$, as mass of the companion star. This results in a value of $q>4.8$, which translates into an upper limit to the orbital period of Swift~J1910.2$-$0546 of 7.4 hrs. However, we note that the amplitude of such modulations from hotspots is expected to be small, and hence not easily detectable during outburst where the optical emission is dominated by the bright, irradiated and/or viscous accretion disk.

Another intriguing possibility is that the observed periodicity of $2.36\pm0.01$ hrs is linked to a superhump modulation instead. Superhumps typically have a slightly longer period than the orbital one, and are caused by tidal stresses in an elliptical disk that is precessing. For this reason, they can only be observed in systems that have a high value of $q$, which is indeed the case for Swift~J1910.2$-$0546. Following \citet{Zurita2008}, we can then obtain an estimate of the orbital period by using the empirical expression derived by \citet{Patterson2005}:

\begin{equation}
\Delta P=0.18 \,(1/q) +0.29\, (1/q)^2,
\end{equation}

where $\Delta P=(P_{\rm sh}-P_{\rm orb})/P_{\rm orb}$ and $P_{\rm sh}$ is the superhump period. Using this equation, we obtain  $\Delta P<5\%$, and therefore $2.25 \, \rm hr < P_{\rm orb} < 2.47 \, \rm hr$. In this scenario, Swift J1910.2--0546 could become the shortest orbital period BHXB known to date \citep[currently, MAXI~J1659$-$152 with $P_{\rm orb} =$ 2.41 hrs is the shortest among the known BHXBs, see][]{kuulkers2011}.

\subsection{Optical / X-ray correlation} \label{sec:optX}

\begin{figure}
\centering
\includegraphics[height=6.6cm,angle=0]{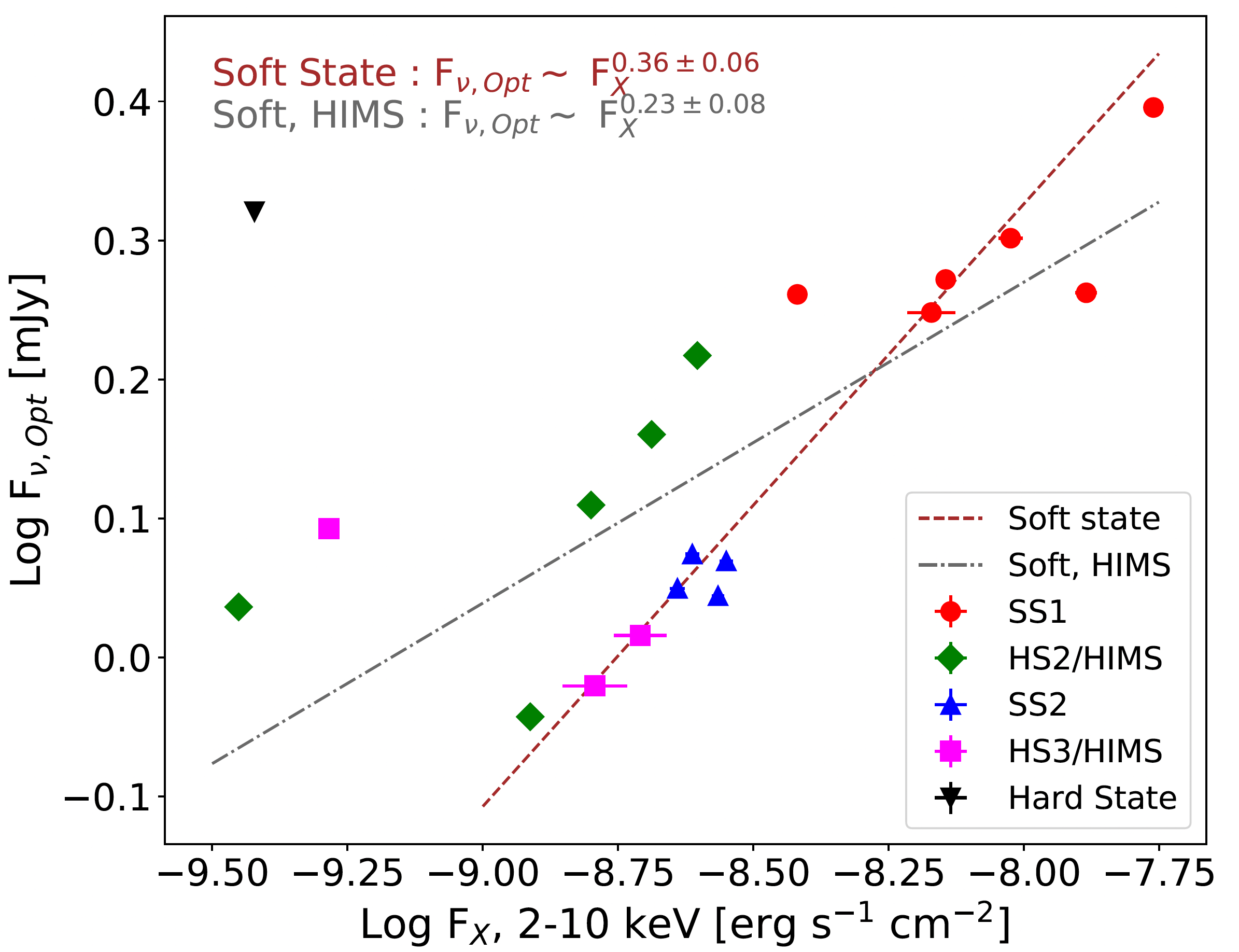}
\caption{Optical/X-ray correlation plot. The different states in the 2012 outburst are plotted in different colors/symbols. The red dashed line shows the best-fit line in the soft state, using the orthogonal distance regression method of least squares. The grey dotted-dashed line is obtained for the complete data including both soft and the HIMS states (excluding the hard state point). While the slope for the soft state only data is $0.36 \pm 0.06$, the correlation is shallower when the HIMS detections are included in the fit, with a slope of $0.23 \pm 0.08$.}
\label{fig:optX-soft}
\end{figure}

The optical/X-ray correlation of Swift~J1910.2$-$0546, plotted with optical data (${i}^{\prime }$-band) from LCO and soft X-ray flux in the 2-10 keV range from \emph{Swift}/XRT, is shown in Fig. \ref{fig:optX-soft}. We plot only quasi-simultaneous data, where the X-ray fluxes are restricted to be within a day of the optical observations. 

The correlation for the soft state data is found to be significant (Pearson correlation coefficient = 0.9, p-value = 0.0002). The power law index of the correlation was found to be $0.36 \pm 0.06$ (where $F_{\nu,Opt} \propto F_{X}^{\beta}$). This is similar to what is generally seen in many BHXBs in the soft state, like GX~339-4 \citep[$\beta =$0.45$\pm$ 0.04,][]{Coriat2009}, 4U~1543-47 \citep[$\beta =$0.13$\pm$ 0.01,][]{Russell2007slope} and XTE~J1550-564 \citep[$\beta =$0.23$\pm$ 0.02,][]{Russell2007slope}. 

Inclusion of both the soft and HIMS data makes the correlation much shallower with $\beta =0.23 \pm 0.08$, although the correlation is still significant (Pearson correlation coefficient = 0.8, p-value = 0.0001). Such shallow slopes were also observed in the UV/X-ray correlation of Swift~J1910.2$-$0546, with $\beta =0.15 \pm 0.05$ \citep{Degenaar2014}. We find that the HIMS data are slightly optically brighter than the soft state data at the same X-ray flux. Moreover the one pure hard state point is clearly way above the soft state and HIMS+soft correlations, possibly following a trend similar to the UV/X-ray correlation, where a significant negative correlation was found at lower fluxes in the hard state \citep{Degenaar2014}. Generally, a range in the value of the slope (or the index) of the power law are observed for the optical/X-ray correlation fits of BHXBs in the hard state, e.g. GS~ 1354$-$64 \citep[$\beta\sim$0.4-0.5,][]{koljonen}, GRS~1716$-$249 \citep[$\beta =0.41\pm$ 0.03,][]{saikia1716}, GX~339$-$4 \citep[$\beta =$0.44$\pm$ 0.01,][]{Coriat2009}, XTE~J1817$-$330 \citep[$\beta =$0.47$\pm$ 0.03,][]{rykoff}, V404~Cyg \citep[$\beta \sim$0.56,][]{Bernardini2016,hynes2019}. However shallow slopes are not unusual and can be seen in other BH and NS LMXBs like Swift~J1357.2$-$0933 \citep[][]{armaspadilla}, SAX~J1808.4$-$3658 \citep{Patruno2016_1808}, Cen~X$-$4 \citep{Baglio_submitted}.

Under simple geometric assumptions, while reprocessed optical emission from an X-ray irradiated accretion disk is expected to be proportional to the X-ray emission with an index of $\sim$0.5 \citep{VanParadijs1994}, a viscously heated disk is expected to have a wavelength-dependent slope around $\sim$0.3 \citep{Russell2006}, and optically thin synchrotron radiation from jets is predicted to result in an even higher correlation index of $\sim$0.7 \citep{corbel2003,Russell2006}. However recent studies with detailed theoretical calculations constrain different ranges of correlation coefficients depending on whether the optical emission is coming from the Rayleigh–Jeans (RJ) tail or closer to the peak of the multicolor blackbody disk \citep[for a detailed calculation, see][]{Coriat2009,Shahbaz2015,be}. They find that for a viscously heated disc, the expected slope can be between 0.13 (RJ) and 0.33 (disk) in the hard state, and 0.26 (RJ) and 0.67 (disk) in the soft state. On the other hand, for X-ray reprocessed disks, the slope changes to a value in the range of 0.14 (RJ) to 0.67 (disk) in the hard state, and 0.28 (RJ) to 1.33 (disk) in the soft state. For Swift~J1910.2$-$0546, the outer disk temperature during outburst reaches $\sim$10000 K \citep{paper2}, where the optical flux is at the spectral transition between the RJ tail and the multicolor blackbody \citep{Russell2006}. Hence based on the slope of the optical/X-ray correlation alone, we cannot confirm whether the optical emission is coming from a viscous disk or a reprocessed disk.

\subsection{Spectral energy distributions} \label{sec:sed}

\begin{table}
\caption{Absorption coefficients for Swift~J1910.2$-$0546 evaluated assuming $N_H=(3.5\pm 0.1)\times 10^{21}\, \rm cm^{-2}$ \citep{Degenaar2014} and the \citep{Foight16} $N_{H}/A_{V}$ relation for our Galaxy. The optical data were de-reddened using the \citet{Cardelli1989} extinction law and the UV data using the \citet{mathis} extinction values.}            
\label{tab_abs_coeff}      
\centering                       
\begin{tabular}{l c l c}       
\hline    
Filter & Central $\lambda$ ($\mu$m)& Instrument & $A_{\lambda}$ (mag)\\
\hline
$K_S$ & 2.147 & LIRIS & $0.14 \pm 0.01$\\
$Y$ & 1.004 & LCO & $0.53 \pm 0.03$\\
${i}^{\prime }$& 0.754 & LCO & $0.81\pm 0.04$\\
$R$ & 0.641 & LCO & $1.03 \pm 0.05$\\
${r}^{\prime }$ & 0.620 & ACAM & $1.07 \pm 0.05$\\
$V$ & 0.545 & LCO & $1.23 \pm 0.06$\\
$B$ & 0.435 & LCO & $1.64 \pm 0.08$\\
$v$ & 0.547 & UVOT& $1.23 \pm 0.06$\\
$b$ & 0.439 & UVOT& $1.62 \pm 0.08$\\
$u$ & 0.346 & UVOT& $1.95 \pm 0.10$\\
$uvw1$& 0.260 & UVOT& $2.62 \pm 0.13$\\
$uvm2$& 0.225 & UVOT& $3.62 \pm 0.18$\\
$uvw2$& 0.193 & UVOT& $3.32 \pm 0.17$\\   
\hline                     
\end{tabular}
\end{table}

\begin{figure}
    \centering
      \includegraphics[height=6.0cm,angle=0]{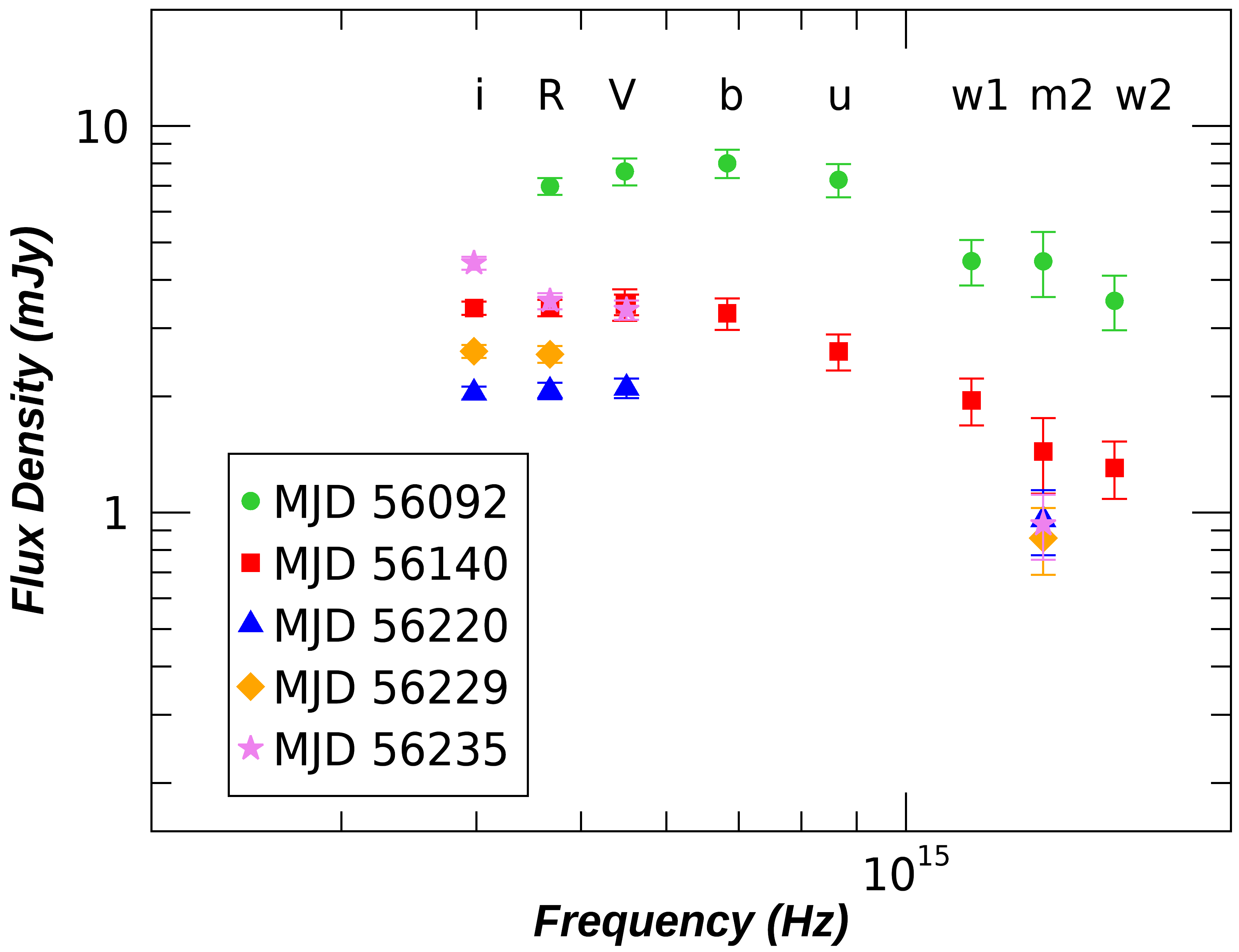}
   \caption{Intrinsic (de-reddened) optical/UV spectra of Swift~J1910.2$-$0546 with quasi-simultaneous data (taken within 24 hours), during selected epochs from the 2012 outburst.}
   \label{Fig:SEDs}
\end{figure}

To understand the various physical processes contributing to the optical emission of Swift~J1910.2$-$0546, we build the de-reddened spectra and spectral energy distributions (SEDs) in some of the most relevant epochs of the 2012 outburst (see Fig. \ref{Fig:SEDs}). 
To build them, we extract the de-reddened fluxes from the calibrated magnitudes using the hydrogen column density ($N_H$) estimate from \citet{Degenaar2014}, $N_H=(3.5\pm0.1)\times 10^{21}\, \rm cm^{-2}$. Using the \citet{Foight16} relation between $N_H$ and the absorption coefficient in $V$-band ($A_{\rm V}$) in our Galaxy, we obtain $A_{\rm V}= 1.22\pm 0.06$. From this, according to the \citet{Cardelli1989} extinction law we obtain the absorption coefficients for the optical and UV filters that are relevant for this work (see Table \ref{tab_abs_coeff}).

We only consider those dates for building the spectra when quasi-simultaneous observations (with data taken within 24 hours) were available in two or more filters. We use the available information to constrain the intrinsic optical/UV spectral index by fitting the function $S_{\nu} \propto \nu^{\alpha}$, where $S_{\nu}$ is the flux density, $\nu$ is the frequency and $\alpha$ is the spectral index. Generally, a spectral index $\sim$1 is expected at optical wavelengths, if the emission is dominated by the blackbody from the outer accretion disk \citep[e.g.][]{hynes2005}, a negative slope $\sim$ -0.7 is expected for optically thin synchrotron emission from the jet \citep[e.g.][]{Gandhi11} and an $\alpha \sim$ 0.3, which turns to a steeper slope 0.3 $< \alpha <$ 2.0 at lower frequencies, is expected for a viscously heated disk \citep[e.g.][]{Accretion_Power}, and an intermediate slope is found in the case when a combination of all the processes contribute towards the optical emission \citep[e.g.][]{saikia1716}.

For Swift~J1910.2$-$0546, we find that the optical/UV spectra are fairly smooth, with a slightly positive (during the relatively brighter epochs) to flat slope (during the fainter epochs) at optical wavelengths, and a negative slope at UV wavelengths. For example, a typical spectrum during the brighter epochs of the  outburst (e.g. MJD 56092 in Fig. \ref{Fig:SEDs}) has a positive optical spectral index $\alpha_{R-b}$ =0.36$\pm$0.26, with a slight peak around the $b$-band, before getting fainter at the higher frequencies, as apparent in the \emph{Swift} UVOT data, with a negative slope ($\alpha_{UV-Opt}$ = -0.99$\pm$0.22). At the fainter epochs of the 2012 outburst, the optical spectra are flatter (eg. $\alpha_{R-b}$ = -0.08$\pm$0.28 on MJD 56140). The ${i}^{\prime }$-band flux is generally found to be brighter than the higher frequencies in the pure HS of 2012 (unlike the SS or the HIMS). The overall shape of the optical/UV spectra is consistent with the outer regions of a blue accretion disk, and is typical for BHXBs in outburst where the optical/UV emissions are dominated by an X-ray irradiated disk.


\begin{figure}
    \centering
    \includegraphics[scale=0.35,angle=270]{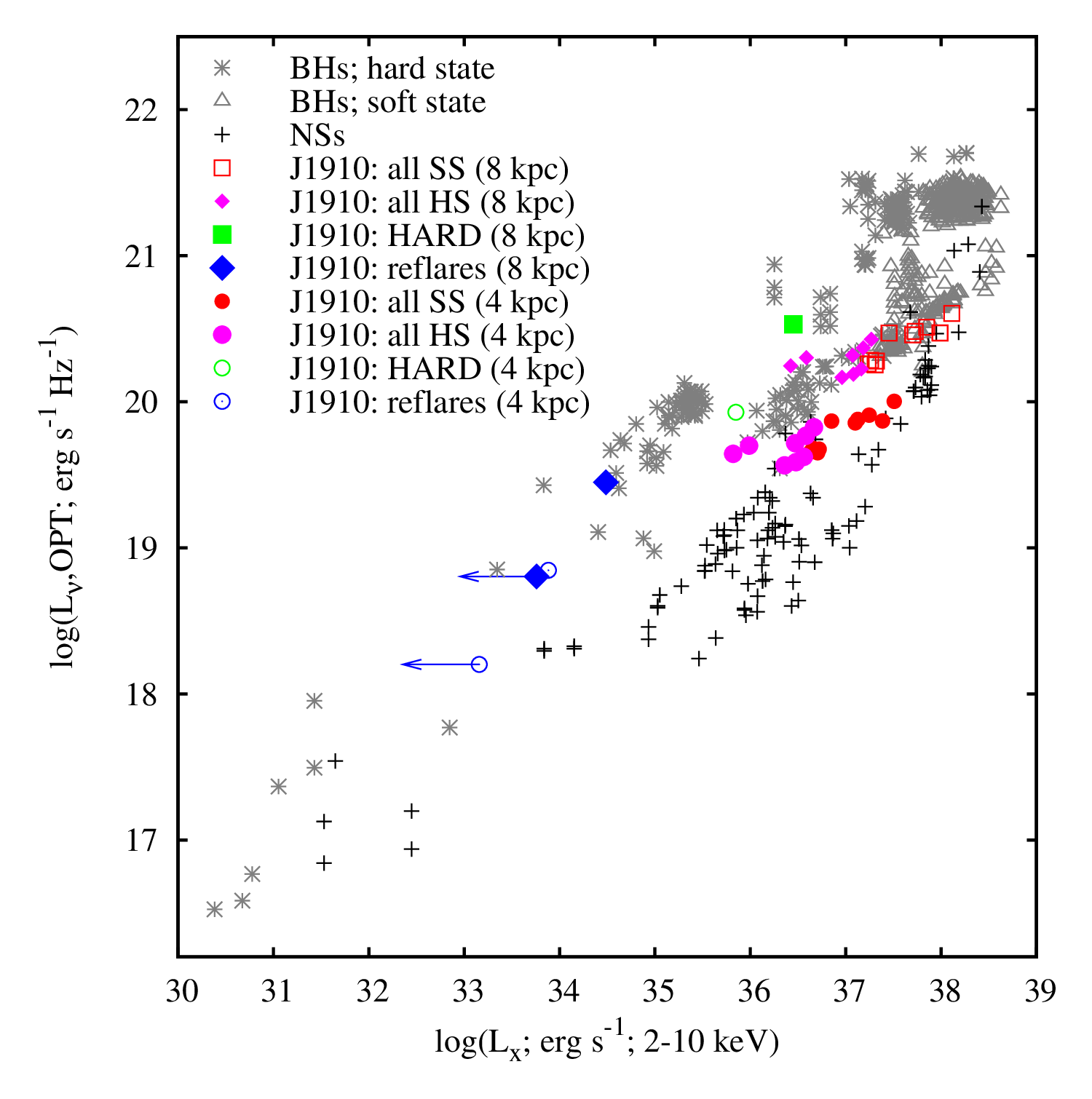}
    \caption{Global, de-reddened, optical/X-ray correlation plot for Swift~J1910.2$-$0546, assuming distances of 4 kpc and 8 kpc (coloured points, as described by the figure legend), in comparison with a sample of BH (grey stars for hard state, and grey open triangles for soft state) and NS (black plus symbols) LMXBs.}
    \label{Fig:optX-global}
\end{figure}

\subsection{Implications on the source distance} \label{sec:dist}

The distance to Swift~J1910.2$-$0546 is not yet confidently known. Previously, \cite{Nakahira14} reported a lower limit of the distance to be $d > 1.7$ kpc. We use two different methods to constrain the distance, using the soft-to-hard state transition luminosity, and the comparative position of the source in the global optical/X-ray correlation.

\subsubsection{Using the state transition luminosity}

BHXBs have been observed to transit from the soft state to the hard state at luminosities between 0.3-3\% percent of the Eddington luminosity \citep{Kalemci2013}, with a mean value of 1.9$\pm$0.2\% \citep{maccarone2003}. Recently, \cite{VahdatMotlagh2019} investigated the distribution of state transition luminosities for a large sample of BHXBs (11 BH sources in 19 different outbursts) from the RXTE archival data, and found that during the transition to the hard state, BHXBs have a mean logarithmic Eddington luminosity fraction of -1.80$\pm$0.25  (as the photon index reaches the lowest value). The state transition luminosity has been used to estimate the distance to many BHXB sources for which a direct measurement of the distance is not available \citep[e.g.][]{homan2006,millerjones2012,saikia1716,williams2022}. For Swift~J1910.2--0546, this transition starts around 2012 Oct 25th (MJD 56225), and the \emph{Swift}/XRT hardness ratio remains at $\sim1.5$ after Nov 10th (MJD 56241.8). During the transition, the unabsorbed X-ray flux (from \emph{Swift}/XRT count-rates) assuming a power law index of 1.7 \citep{Degenaar2014}, is found to be 2.57$\times10^{-10}$ ergs/cm$^2$/s in the 2-10 keV range and 1.38$\times10^{-9}$ ergs/cm$^2$/s in the 0.5-200 keV bolometric range. The mass of the BH is not confirmed, while \cite{Nakahira14} provided a lower limit for the mass of the compact object as $M_{\rm BH} > 2.9 M_{\odot}$, recently \cite{new} constrained a mass range of $M_{\rm BH} = 6.31$–$13.65 M_{\odot}$. Assuming a black hole mass of 8$\pm$1.5 $M_{\odot}$ \citep[mean value for BHXBs with uncertain masses,][]{VahdatMotlagh2019}, we infer the distance to the source to be in the range of 6.7---14.3 kpc, similar to the result obtained by \cite{new}. For a more conservative BH mass range of 3--20 $M_{\odot}$, the distance to the source can be expected to be in the range of 4.5--20.8 kpc.

\subsubsection{Using the global optical/X-ray correlation}

An independent method of constraining the source distance, using the global optical/X-ray correlation of the source with respect to a sample of other BH and NS LMXBs \citep{Russell2006, Russell2007}, also supports this finding. Generally, both classes of LMXBs have different correlation tracks in the optical/X-ray correlation plot, with the BH LMXBs being optically brighter than the NS LMXBs \citep{Russell2006,Bernardini2016}. This comparative study is helpful in constraining the distance to a source, when a direct measurement of the distance is not available \citep[see e.g. the cases of BHXBs 4U~1957+11 and GRS~1716$-$249;][]{Russell20104u,saikia1716}. We plot in Fig. \ref{Fig:optX-global}, the comparative de-reddened optical/X-ray correlation of Swift~J1910.2$-$0546 with respect to other BH and NS LMXBs, assuming distances of 4 kpc and 8 kpc. In addition to the 2012 outburst data, we also include in the plot quasi-simultaneous optical and X-ray magnitudes acquired during the reflares of Swift~J1910.2$-$0546 in 2013 \citep{paper2}. We find that the reflare data (shown in green squares) lie clearly within the hard state BH LMXBs, which supports the assumption that the source was in the hard state during those reflares, and explains why Swift/XRT did not detect the source in the second observation, since the source is still consistent with the hard state correlation and would be too faint to be detected with Swift. We find that the location of the source on the optical/X-ray luminosity correlation is typical for a BHXB if the distance to the source is $\gtrsim 8$ kpc, as a shorter distance pushes the source to the neutron star regime. However we note that although in the historical data set \citep{Russell2006,Russell2007} there are no soft state data below $L_{X}\sim10^{37.3}$, there have been recent BHXBs like Swift~J1753.5--0127 which have soft state data at lower luminosities. Moreover, there are some NS Z-sources that overlap with soft state BHs at very high luminosities. Hence it is not straightforward to use soft state data to constrain the distance using this method. So from our comparative analysis, it is difficult to constrain a distance range, but we find that a fairly near distance would imply a low luminosity soft state, which is unusual and agrees with Section 3.7.1.

\section{Summary and conclusions} \label{sec:conclusions}

In this work, we present multi-wavelength monitoring of the candidate black hole transient X-ray binary Swift~J1910.2$-$0546 during its 2012 outburst, mainly focusing on optical data taken with the Faulkes telescopes, presented here for the first time. We observe a prominent dip in the optical intensity a few days before the observed dip in UV and X-ray wavelengths, probably caused by radial propagation of mass instability from the outer edge of the accretion disk towards the inner region. Just after the dip, Swift~J1910.2$-$0546 exhibits another state transition, where the soft-to-hard transition occurred at a higher luminosity than the hard-to-soft transition. This results in a highly unusual clockwise loop in the hardness--intensity diagram, which could be due to brightening of the disk after the dip in the intensity while the power law stays relatively unchanged. 

The spectral energy distributions suggest that the optical/UV emission of the source is mostly dominated by an X-ray irradiated accretion disk. However, when the source transitions into the pure hard state, we observe a dramatic change in the optical colour, and an optical brightening most likely due to the onset of the jet during the transition. From our high cadence observations taken during three different epochs, we find significant variability and a hint of a sinusoidal modulation in the optical emission. From the periodicity of the modulation, we constrain an upper limit of 7.4 hrs to the orbital period of Swift~J1910.2$-$0546 under the assumption that the modulation arises in a hot spot in the outer disk. If instead it is a superhump modulation, then the orbital period is likely to be 2.25--2.47 hr, making Swift~J1910.2$-$0546 the shortest orbital period BHXB ever known. The slope of the optical/X-ray correlation of Swift~J1910.2$-$0546 is found to be shallow. Comparing the soft and the HIMS data of Swift~J1910.2$-$0546 against a global sample of other BH and NS LMXBs, we find that the correlation is typical for a BHXB if the distance to the source is $\gtrsim 8$ kpc. We show that this is also supported by the state transition luminosity of the source, which suggests a distance of 4.5--20.8 kpc to Swift~J1910.2$-$0546.

\section*{Acknowledgements}

The authors thank the anonymous referee for useful comments and suggestions. DMR and DMB acknowledge the support of the NYU Abu Dhabi Research Enhancement Fund under grant RE124. This work uses data from the Faulkes Telescope Project, which is an education partner of Las Cumbres Observatory (LCO). The Faulkes Telescopes are maintained and operated by LCO. This work also makes use of data supplied by the UK \emph{Swift} Science Data Centre at the University of Leicester, and the MAXI data provided by RIKEN, JAXA and the MAXI team.

\section*{Data Availability}

The optical data from Faulkes/LCO underlying this article will be shared on reasonable request to the corresponding author. All the Swift data are publicly accessible through the HEASARC portal (https://heasarc.gsfc.nasa.gov/db-perl/W3Browse/w3browse.pl) and the UK Swift Science Data Centre (https://www.Swift.ac.uk/archive/index.php). The MAXI data underlying this article are publicly available at http://maxi.riken.jp/top/index.html.



\bibliographystyle{mnras}
\bibliography{bib} 





\bsp	
\label{lastpage}
\end{document}